\documentclass[twocolumn,10pt]{IEEEtran}

\usepackage{graphicx}
\usepackage{amsmath}
\usepackage{cite}
\usepackage{multirow}
\usepackage{subfigure}

\usepackage{amssymb}
\usepackage{color}

\begin{document}

\title{High-Mobility OFDM Downlink Transmission with Large-Scale Antenna Array}

\author{Wei Guo, Weile Zhang, Pengcheng Mu and Feifei Gao, {\it Senior Member, IEEE}
\thanks{\hspace{-0.0cm}W. Guo, W. Zhang and P. Mu are with the School of Electronic and Information Engineering, Xi'an Jiaotong University, Xi'an, Shaanxi, 710049, China (e-mail: guowei@stu.xjtu.edu.cn, wlzhang@mail.xjtu.edu.cn, pcmu@mail.xjtu.edu.cn).

F. Gao is with Tsinghua National Laboratory for Information Science and Technology (TNList), Beijing, 100084, China (e-mail: feifeigao@ieee.org).
}}
\maketitle

\begin{abstract}
In this correspondence, we propose a new receiver design for high-mobility orthogonal frequency division multiplexing (OFDM) downlink transmissions with a large-scale antenna array. The downlink signal experiences the challenging fast time-varying propagation channel. The time-varying nature originates from the multiple carrier frequency offsets (CFOs) due to the transceiver oscillator frequency offset (OFO) and multiple Doppler shifts.
Let the received signal first go through a carefully designed beamforming network, which could separate multiple CFOs in the spatial domain with sufficient number of receive antennas. A joint estimation method for the Doppler shifts and the OFO is further developed. Then the conventional single-CFO compensation and channel estimation method can be carried out for each beamforming branch. The proposed receiver design avoids the complicated time-varying channel estimation, which differs a lot from the conventional methods. More importantly, the proposed scheme can be applied to the commonly used time-varying channel models, such as the Jakes' channel model.
\end{abstract}


\section{Introduction}
Orthogonal frequency division multiplexing (OFDM) has gained much interests over the last couple of decades since it can effectively deal with frequency-selective channels~\cite{Abdzadeh11Robust}. However, an OFDM system is sensitive to the carrier frequency offsets (CFOs). There are two kinds of CFOs, one is caused by the mismatch of local oscillators, and the other is incurred by the relative moving between the transmitter and receiver. They are called oscillator frequency offset (OFO) and Doppler frequency offset (DFO), respectively, in the literature \cite{Abdzadeh11Robust,Schmidl97Robust,Morelli00Carrier,Souden09Robust,Nguyen10Pilot}. The presence of CFOs makes the channel vary with time rapidly and thus introduces inter-carrier interference (ICI) that could significantly degrade the link performance.

The single CFO, such as the OFO, is easy to deal with using some previous methods~\cite{Abdzadeh11Robust,Schmidl97Robust,Morelli00Carrier}. However, when the receiver or the transmitter moves at a high speed, the effect of the various Doppler shifts associated with various multipaths becomes very serious~\cite{Souden09Robust,Nguyen10Pilot}. In this case, the existence of multiple DFOs accompanying single OFO greatly increases the complexity of CFOs estimation and compensation. In~\cite{Souden09Robust}, the maximum Doppler spread and the OFO are jointly estimated by covariance-matching which requires the statistical information of the channel. With the knowledge of the CFOs, the time-varying channel estimation is still a challenge due to the vast number of parameters to be estimated. The basic expansion model (BEM) has been used to characterize the time-variations of channel with reduced number of parameters~\cite{Nguyen10Pilot}.
In~\cite{Nguyen10Pilot}, a maximum likelihood (ML) based estimator for the OFO and the BEM coefficients is exploited with the accurate knowledge of the maximum Doppler shift.

Different from the OFO, the multiple DFOs are associated with the angle of arrivals (AoAs) for multipaths. Some previous work has been done to mitigate the effect of the DFOs from spatial domain via small-scale antenna arrays~\cite{Zhang11Multiple,Yang13Beamforming}. In~\cite{Zhang11Multiple}, the DFOs associated with the dominant multipaths in a sparse channel are compensated separately with the knowledge of the maximum Doppler shift and the AoAs. Similarly, the DFOs for the line-of-sight (LoS) from different base stations (BSs) are also compensated via AoAs estimation in~\cite{Yang13Beamforming}.
Most of these works are based on sparse channel with limited multipaths. When there are a large number of multipaths from the BS to the moving terminal caused by the rich reflectors, it is hard to get the AoAs for all paths and separate the multipaths with a conventional small-scale antenna array.

Fortunately, large-scale antenna array can provide high-spatial resolution~\cite{Xie16An,Xie17Unified} and may facilitate the resistance of the AoA-related multiple Doppler shifts in high-mobility communications. Note that the large-scale antenna array technique, known as ``massive MIMO'', has drawn considerable interests from both academia and industry more recently.
Motivated by the above observation, in this correspondence we propose a new receiver design for OFDM downlink transmissions with a large-scale antenna array.
We first separate multiple CFOs in the spatial domain through a carefully designed beamforming network, and a joint estimation method for the Doppler shifts and the OFO is further developed. Then the conventional single-CFO compensation and channel estimation method can be carried out for each beamforming branch.
The proposed scheme can be applied to the commonly used time-varying channel models, such as the Jakes' channel model. Simulation results also show the effectiveness of the proposed scheme.

\emph{Notations}: $(\cdot)^\mathrm{H}$, $(\cdot)^\mathrm{T}$, and $(\cdot)^\ast$ represent the Hermitian, transposition, and conjugate operations, respectively. $\Re(\cdot)$, $\Im(\cdot)$, and $\|\cdot\|$ denote the real part, imaginary part, and Euclidean norm of a vector, respectively. %$|\cdot|$ denotes the absolute value of a scalar

\vspace*{-0.0in}
\section{System Model}\label{sec:SysMod}
\begin{figure}[t!]
\centering
\includegraphics[scale=0.35]{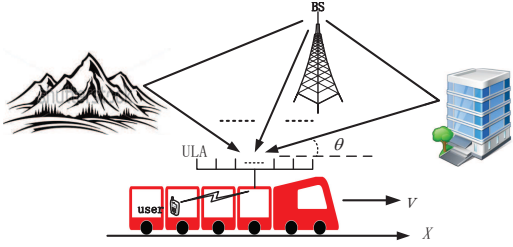}
\caption{Downlink transmission with high-mobility terminals.}
\label{fig:SysMod}
\vspace*{-0.0in}
\end{figure}

We consider the scenarios of high-mobility downlink communications. As in Figure~\ref{fig:SysMod}, the BS transmits signal to the users in a high-speed train through multipaths, which are caused by the various reflectors, such as the buildings or mountains. A relay station with a large-scale uniform linear array (ULA) is configured on top of the train for decoding and forwarding the data between the BS and the users~\cite{Zhang11Multiple}. The direction of the ULA agrees with that of the motion.

\vspace*{-0.0in}
\subsection{Transmitted Signal Model}\label{ssec:TxSigMod}
Consider the frame structure in an OFDM system, where each frame consists of $N_b$ OFDM blocks. The first block is the training block which is known at the transmitter and receiver for timing, CFOs estimation, and channel estimation. The training block contains two identical halves in time domain. Note that this structure has been widely used in the literature for OFO estimation~\cite{Schmidl97Robust}. The remaining blocks are used for transmitting data symbols to the receivers.

Denote $N_c$ as the number of subcarriers. After applying an $N_c$-point inverse discrete Fourier transform (IDFT) operator to the information symbols and adding the cyclic prefix (CP) of length $N_{cp}$ to each block, the resulting time-domain samples in the $m$th block are given by $s_m(n),-N_{cp}\leq n\leq N_{c}-1$.
Let $N_s=N_c+N_{cp}$ be the length of an OFDM block. Then the time-domain OFDM transmitted signal can be written as
\begin{equation}\label{eq:s_n}
  \tilde{s}\left(n\right)=\sum\limits_{m=0}^{\infty}{s_m\left( n-m{N_s} \right)}.
\end{equation}

\vspace*{-0.0in}
\subsection{Time-Varying Channel Model}\label{ssec:ChnMod}
As in Figure~\ref{fig:SysMod}, we consider the downlink transmission between the BS and the high-speed train, and there are rich reflectors around the moving train. Assume that the large-scale ULA has $N_r$ antennas, then the baseband time-varying multipath channel from the BS to the $a$th antenna can be modeled as $L$ taps with different delays,
\begin{equation}\label{eq:h_chn}
  h_a\left(n,n'\right) = \sum\limits_{l=1}^{L}{g_{a,l}(n)\delta\left(n'-d_l\right)},
\end{equation}
where $d_l$ is the relative delay of the $l$th tap, and $g_{a,l}(n)$ is the corresponding complex amplitude for the $l$th tap at the $a$th antenna. %In a high-mobility environment, $g_{a,l}(n)$ varies with the time index $n$ due to the significant Doppler spread.
In~\cite{Zheng03Simulation}, an established Jakes' channel model simulator has been proposed to characterize the scenarios with rich reflectors. As in~\cite{Zheng03Simulation}, each tap is modeled as a frequency-nonselective fading channel comprised of $N_p$ propagation paths. The equivalent model for $g_{a,l}(n)$ can be expressed as
\begin{equation}\label{eq:h_tap}
  g_{a,l}(n) = E_l \sum\limits_{q=1}^{N_p} {\alpha_{l,q} e^{j[2\pi{f_d}n{T_s}\cos{\theta_{l,q}}+\phi_a(\theta_{l,q})]} },
\end{equation}
where $E_l$ is a scaling constant for the $l$th tap, $\alpha_{l,q}$ is the random complex path gain associated with the $q$th propagation path in the $l$th tap, and $T_s$ is the sampling interval. The maximum Doppler shift, $f_d$, is defined as $f_d=v/\lambda$, where $v$ is the speed of the moving terminal, and $\lambda$ is the wavelength of carrier wave. $\theta_{l,q}$ is the AoA of the $q$th path in the $l$th tap relative to the direction of motion, and it is random distributed between 0 and $\pi$ in the three-dimensional (3D) space. Thus the DFO for the $q$th path in the $l$th tap is determined by $f_{l,q}={f_d}\cos{\theta_{l,q}}$.
The phase shift at the $a$th antenna is denoted as $\phi_a(\theta_{l,q})$ in~\eqref{eq:h_tap}. It is determined by the antenna structure, position, and the AoA, $\theta_{l,q}$. Take the 1st antenna as a reference, the phase shift, $\phi_a(\theta_{l,q})$, is given by
\begin{equation}
  \phi_a(\theta_{l,q})=2\pi(a-1)d\cos\theta_{l,q}/\lambda,
\end{equation}
where $d<\lambda/2$ is the antenna element spacing. We can further determine the steering vector for the whole antenna array at the direction $\theta_{l,q}$ as $\mathbf{a}(\theta_{l,q})=[e^{j\phi_1(\theta_{l,q})},\cdots,e^{j\phi_{N_r}(\theta_{l,q})}]^\mathrm{T}$.

In a high-mobility environment, $g_{a,l}(n)$ varies with the time index $n$ due to the significant Doppler shifts.
Each path in~\eqref{eq:h_tap} has independent attenuation, phase, AoA and also DFO. During the transmitting period of one OFDM frame, there is little change in the position and speed of the moving terminal. Therefore, we can assume that $\alpha_{l,q}$, $\theta_{l,q}$, and $f_d$ are constant over the observed data frame, and only vary among different frames.
When there are rich scatters around the moving terminal, $N_p$ tends to be vary large to reflect the classical Jakes' channel model~\cite{Jakes94Microwave}, while when there are few dominant multipaths from the BS to the moving terminal,~\eqref{eq:h_chn} is simplified to the sparse channel model in~\cite{Zhang11Multiple}. Thus, the channel model discussed in this correspondence reflects various scenarios of the practical high-mobility transmission.

\vspace*{-0.0in}
\subsection{Received Signal Model}\label{ssec:RxSigMod}
After passing over the above time-varying multipath channel, the received signal at the $a$th antenna is given by
\begin{align}\label{eq:y_a}
  \tilde{y}_a(n) = \sum\limits_{l=1}^{L} g_{a,l}(n) e^{j2\pi{\varepsilon}n{T_s}} \tilde{s}(n-d_l) + \tilde{z}_a(n),
\end{align}
where $\varepsilon$ is the OFO between the transceivers, and $\tilde{z}_a(n)$ is the complex additive white Gaussian noise at the $a$th receive antenna with mean zero and variance $\sigma_n^2$.

Assume perfect time synchronization at the receiver. From~\eqref{eq:s_n},~\eqref{eq:h_tap}, and~\eqref{eq:y_a}, the $n$th time-domain sample in the $m$th OFDM block at the $a$th antenna can be expressed as
\begin{align}\label{eq:y_amn}
  y_{a,m,n} & = \tilde{y}_a\left(m{N_s}+n\right) \nonumber\\
  & \!=\! \sum\limits_{l=1}^{L} \sum\limits_{q=1}^{N_p} e^{j\phi_a(\theta_{l,q})} \beta_{m,n}(\theta_{l,q}) \hat{s}_{m,n}(d_l) \!+\! z_{a,m,n},
\end{align}
where $n=0,1,\cdots ,N_c-1$ is the time index in a block, $\beta_{m,n}(\theta_{l,q})$ includes the phase rotation introduced by both the DFO, $f_{l,q}$, and the OFO, $\varepsilon$, and is expressed as $\beta_{m,n}(\theta_{l,q})={E_l}\alpha_{l,q} e^{j2\pi\hat{f}_{l,q}(m{N_s}+n){T_s}}$, where $\hat{f}_{l,q}=f_{l,q}+\varepsilon$ is the total CFO for the $q$th path in the $l$th tap, $\hat{s}_{m,n}(d_l)=s_m(n-d_l)$ is the $n$th time-domain sample of the transmitted signal in the $m$th block after the delay of $d_l$, and $z_{a,m,n}=\tilde{z}_a(m{N_s}+n)$ is the $n$th time-domain sample of noise in the $m$th OFDM block at the $a$th receive antenna.

Denote $\mathbf{y}_{m,n} = [y_{1,m,n},\cdots,y_{N_r,m,n}]^\mathrm{T}$ and $\mathbf{z}_{m,n}=[z_{1,m,n},\cdots,z_{N_r,m,n}]^\mathrm{T}$ as the vector representation of the $n$th time-domain sample of the received signal and noise, respectively, in the $m$th OFDM block on the whole antenna array. From~\eqref{eq:y_amn}, we can obtain the received signal matrix for the $m$th OFDM block on the whole antenna array as
\begin{equation}\label{eq:y_m}
  \mathbf{Y}_{m} = \sum\limits_{l=1}^{L} \sum\limits_{q=1}^{N_p} \mathbf{a}(\theta_{l,q}) \mathbf{v}_m^\mathrm{T}(\theta_{l,q}) \mathbf{S}_m(d_l) + \mathbf{Z}_{m},
\end{equation}
where $\mathbf{Y}_{m} = [\mathbf{y}_{m,0},\cdots,\mathbf{y}_{m,N_c-1}] \in \mathbb{C}^{N_r\times N_c}$, $\mathbf{v}_m(\theta_{l,q}) = [\beta_{m,0}(\theta_{l,q}),\cdots,\beta_{m,N_c-1}(\theta_{l,q})]^\mathrm{T} \in \mathbb{C}^{N_c \times 1}$, $\mathbf{S}_{m}(d_l) = \mathrm{diag}[\hat{s}_{m,0}(d_l),\cdots,\hat{s}_{m,N_c-1}(d_l)] \in \mathbb{C}^{N_c\times N_c}$, and $\mathbf{Z}_{m}=[\mathbf{z}_{m,0},\cdots,\mathbf{z}_{m,N_c-1}] \in \mathbb{C}^{N_r\times N_c}$.

We can see from~\eqref{eq:y_amn} that the received signal is affected by multiple CFOs due to one unique OFO and multiple DFOs. If there is no Doppler spread, the time variation is induced by single OFO as in~\cite{Schmidl97Robust,Morelli00Carrier}.
In this case, the single CFO can be directly estimated and compensated. However, the multipath channel with Doppler shifts makes it hard to distinguish multiple CFOs which are mixed at the receiver.

\vspace*{-0.0in}
\section{Receiver Design for High-Mobility OFDM Systems}\label{sec:receiver}
\begin{figure}[t!]
\centering
\includegraphics[scale=0.5]{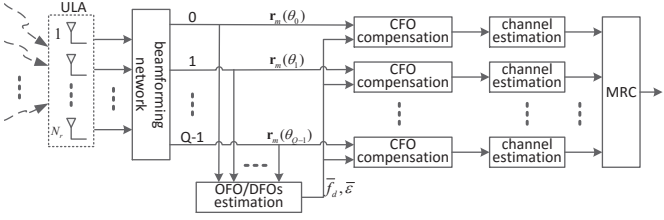}
\caption{Diagram of the receiver design.}
\label{fig:Receiver}
\vspace*{-0.0in}
\end{figure}
In this section, we develop the design procedures at the receiver. Figure~\ref{fig:Receiver} shows the diagram of the receiver design. The received signal first goes through a high-resolution beamforming network which utilizes the baseband digital beamforming technique to produce $Q$ parallel beamforming branches. Then, the Doppler shifts and the OFO are jointly estimated. Since each beamforming branch is mainly affected by one CFO, it is easy to perform CFO compensation and channel estimation. Finally, $Q$ parallel beamforming branches are combined together to recover the transmitted signal.

\vspace*{-0.0in}
\subsection{Beamforming Network}\label{ssec:bf}
As discussed previously, the received signal on a single antenna consists of multiple CFOs due to multiple DFOs and one unique OFO. It is hard to separate multiple CFOs with one single antenna. Since the DFOs are associated with the AoAs, we will solve this problem in spatial domain. Recently, large-scale antenna systems have gained much interest~\cite{Xie16An,Xie17Unified},  where the transmitter or the receiver are equipped with large number of antennas.
These systems can provide high-spatial resolution which is sufficient for dealing with large number of DFOs with different AoAs. In the proposed scheme, we configure a large-scale ULA at the receiver to separate multiple CFOs via beamforming.

To make the received signal contain single DFO, the goal of beamforming is to maintain the signal from only one desired direction while suppressing the signals from other directions \cite{Choi00A}. It can be easily implemented through the matched filter beamformer. The beamforming weight vector for direction $\theta$ is determined by the steering vector, that is,
\begin{equation}\label{eq:w}
  \mathbf{w}(\theta) = \mathbf{a}(\theta)/{N_r}.
\end{equation}

Then the received signal for the $m$th OFDM block after beamforming towards the direction $\theta$ is given by
\begin{equation}\label{eq:r_m}
  \mathbf{r}_m(\theta) = \mathbf{w}^\mathrm{H}(\theta) \mathbf{Y}_{m},
\end{equation}
where $\mathbf{r}_m(\theta) = [r_{m,0}(\theta),\cdots,r_{m,N_c-1}(\theta)]$. From~\eqref{eq:y_m} and~\eqref{eq:w}, we can further divide~\eqref{eq:r_m} into
\begingroup\makeatletter\def\f@size{9}\check@mathfonts
\def\maketag@@@#1{\hbox{\m@th\normalsize\normalfont#1}}%
\begin{align}\label{eq:r_m_1}
  &\mathbf{r}_m(\theta)= \underbrace{\sum\limits_{l,q,\theta_{l,q}=\theta}\mathbf{v}_m^\mathrm{T}(\theta_{l,q}) \mathbf{S}_m(d_l)}_{desired\ signal} \nonumber\\
  &+\!\!\!\! \underbrace{\sum\limits_{l',q',\theta_{l',q'}\neq\theta}\!\!\!\! \mathbf{w}^\mathrm{H}(\theta) \mathbf{a}(\theta_{l',q'}) \mathbf{v}_m^\mathrm{T}(\theta_{l',q'}) \mathbf{S}_m(d_{l'})}_{interference} + \underbrace{\mathbf{w}^\mathrm{H}(\theta)\mathbf{Z}_{m}}_{noise},
\end{align}\endgroup
where the first term is the desired signal from $\theta$ which is the same as the standard received signal model with single CFO. The second term is the interference from other directions which is greatly suppressed through the high-resolution beamforming techniques. The third term is the noise after beamforming. When there is no path from $\theta$, the resultant signal in~\eqref{eq:r_m_1} is comprised of weak interference and noise.

\begin{figure}[t!]
    \centering
    \subfigure[]{\label{fig:BP_1}\includegraphics[scale=0.37]{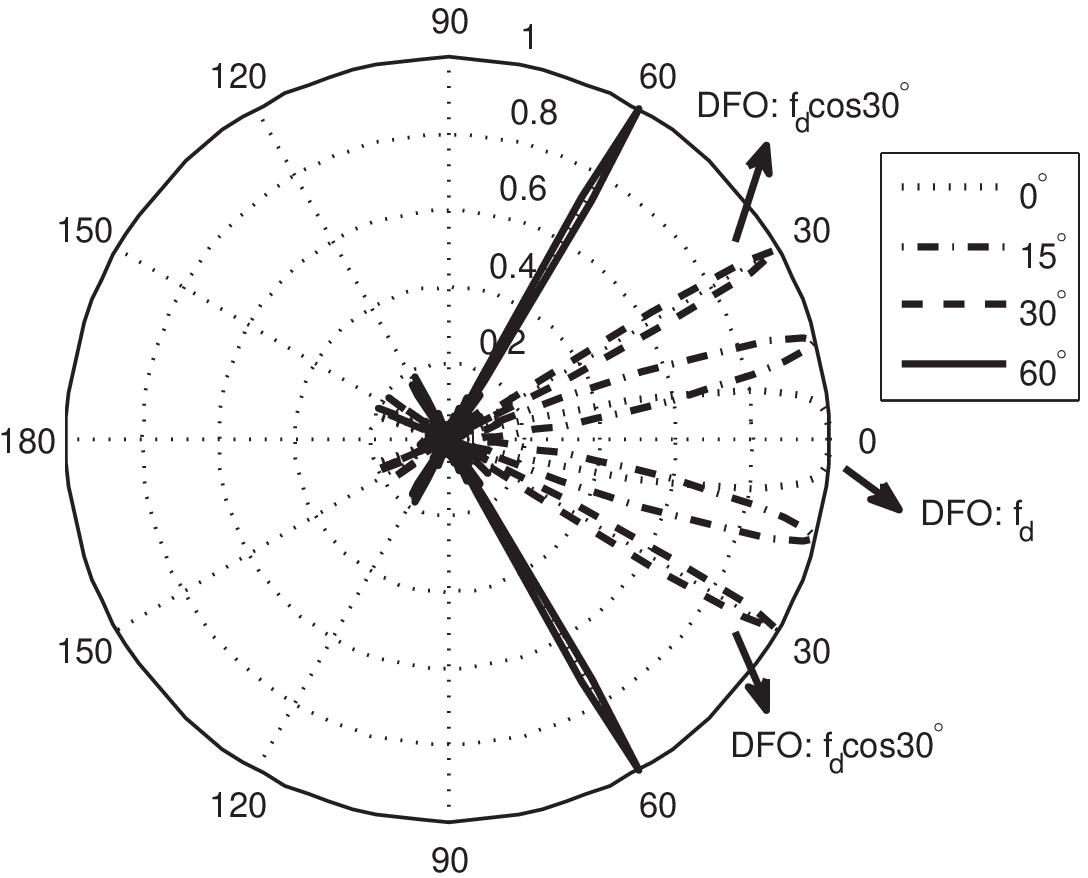}}
    \subfigure[]{\label{fig:BP_2}\includegraphics[scale=0.5]{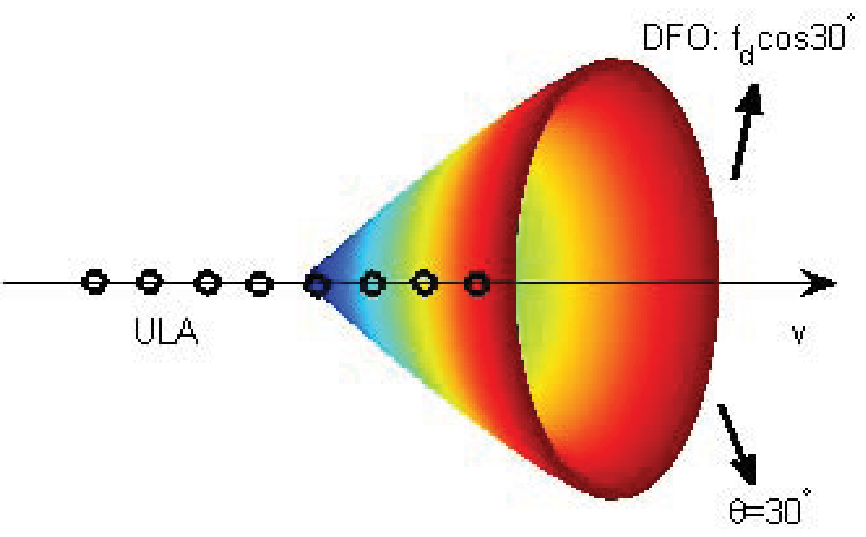}}
    \caption{Beam patterns for a ULA at different AoAs, $d=0.45\lambda$: (a) Two-dimensional (2D) perspective; (b) 3D perspective.}
    \label{fig:Beampattern}
    \vspace*{-0.0in}
\end{figure}

With the weight vectors in~\eqref{eq:w}, Figure~\ref{fig:Beampattern} shows an example of the beam patterns for a ULA under different values of $\theta$. From this figure, the following observations can be made:
\begin{itemize}
\item The beam pattern is symmetric relative to the axial direction of the ULA. Since the DFO is determined by the AoA, the distribution of the DFOs is also symmetric relative to the direction of motion. Thus, the ULA placed in the moving direction can make sure that the received signal in~\eqref{eq:r_m_1} is mainly affected by one significant DFO;
\item The beam width of the main lobe gets wider when $\theta$ gets closer to $0^\circ$ or $180^\circ$, and the DFO varies slowly around $\theta=0^\circ$ and $\theta=180^\circ$, which is determined by the first-order derivative of $f_d\cos\theta$, that is, $-f_d\sin\theta$. Therefore, though the beam width around $0^\circ$ or $180^\circ$ is slightly wide, the received signal in~\eqref{eq:r_m_1} has a narrow Doppler spread which is regarded as one equivalent DFO.
\end{itemize}

Through the above analysis, we place the large-scale ULA in the direction of the motion with $d<\lambda/2$. Moreover, the high-speed train is usually with a long length, it provides sufficient space for deploying the large-scale ULA.
To cover all the possible AoAs of the multipaths, the range for beamforming is designed between $0^\circ$ and $180^\circ$. We further assume that $Q$ different beamforming angles, denoted as $\theta_i$, $i=0,1,\cdots,Q-1$, are evenly selected with the interval of $\Delta$ degree, then $Q\!=\!\lfloor 180 /\Delta\rfloor+1$. %$Q$ different values of $\theta$ are evenly selected between $0^\circ$ and $180^\circ$ to design the beamforming network.
Note that the beamforming is performed towards the pre-selected angles simultaneously, and the beamforming network is designed previously. The AoAs of the multipaths is not required in the proposed method. This is quite different from the existing works~\cite{Zhang11Multiple,Yang13Beamforming}.

\vspace*{-0.0in}
\subsection{Joint Doppler Shifts and OFO Estimation}\label{ssec:fo_est}
After the high-resolution beamforming in Section~\ref{ssec:bf}, the multiple CFOs are separated in spatial domain, and each beamforming branch is mainly affected by one CFO which is the sum of single DFO and single OFO. Now we will estimate the CFOs for $Q$ parallel branches. Since the beamforming angles are selected previously, the CFO for each branch is determined by the maximum Doppler shift, $f_d$, and the OFO, $\varepsilon$, which are invariant in $Q$ parallel branches. Thus, we will jointly estimate the maximum Doppler shift, $f_d$, and the OFO, $\varepsilon$, through the set of beamforming branches instead of estimating the CFO for each branch separately.
Assume perfect interference elimination through large-scale ULA, then the second term in~\eqref{eq:r_m_1} is first neglected.
As in~\cite{Schmidl97Robust}, the correlation between the first and second half of the training block in the $i$th beamforming branch can be computed as
\begin{align}
  b(\theta_i) &= \frac{\sqrt{N_c}}{\|\mathbf{r}_{0}(\theta_i)\|} \sum\limits_{n=0}^{N_c/2-1} r_{0,n}^{*}(\theta_i)r_{0,n+N_c/2}(\theta_i) \nonumber\\
  & = {\rho_i}e^{j2\pi(f_d\cos{\theta_i}+\varepsilon)N_c/2{T_s}}+\eta_i,
\end{align}
where $i=0,1,\cdots,Q-1$, $\rho_i$ is the real coefficient for the $i$th branch, and $\eta_i$ is the noise term in the $i$th branch which is normalized by the coefficient, ${\sqrt{N_c}}/{\|\mathbf{r}_{0}(\theta_i)\|}$. Now the optimization problem can be formulated as
\begin{equation}\label{eq:opt_1}
  \min_{\tilde{\mathbf{\Gamma}},\tilde{f}_d,\tilde{\varepsilon}} \left\|\mathbf{E}(\tilde{f}_d)\tilde{\mathbf{\Gamma}} e^{j\pi\tilde{\varepsilon} N_c{T_s}}-\mathbf{b}\right\|^2,
\end{equation}
where $\mathbf{E}(\tilde{f}_d)=\mathrm{diag}(e^{j\pi \tilde{f}_d\cos{\theta_0}N_c{T_s}},\cdots,e^{j\pi \tilde{f}_d\cos{\theta_{Q-1}}N_c{T_s}})$, $\tilde{\mathbf{\Gamma}}=[\tilde{\rho}_0,\cdots,\tilde{\rho}_{Q-1}]^\mathrm{T}$, $\mathbf{b}=[b(\theta_0),\cdots,b(\theta_{Q-1})]^\mathrm{T}$, and $\tilde{\rho}_i$, $\tilde{f}_d$, and $\tilde{\varepsilon}$ are the unknown real coefficient, maximum Doppler shift and OFO, respectively, that need to be optimized.
To solve this problem, we first reformulate~\eqref{eq:opt_1} as
\begin{equation}\label{eq:opt_2}
  \min_{\tilde{\mathbf{\Gamma}},\tilde{f}_d,\tilde{\varepsilon}} \left\|\tilde{\mathbf{\Gamma}} - \mathbf{E}^\mathrm{H}(\tilde{f}_d)\mathbf{b}e^{-j\pi\tilde{\varepsilon} N_c{T_s}}\right\|^2.
\end{equation}

Since $\tilde{\mathbf{\Gamma}}$ is a vector with real elements,~\eqref{eq:opt_2} is minimized when $\tilde{\mathbf{\Gamma}} = \Re[\mathbf{E}^\mathrm{H}(\tilde{f}_d)\mathbf{b}e^{-j\pi\tilde{\varepsilon} N_c{T_s}}]$, then~\eqref{eq:opt_2} turns to
\begin{equation}\label{eq:opt_3}
  \min_{\tilde{f}_d,\tilde{\varepsilon}} \left\|\Im[\mathbf{E}^\mathrm{H}(\tilde{f}_d)\mathbf{b}e^{-j\pi\tilde{\varepsilon} N_c{T_s}}]\right\|^2.
\end{equation}

Since $\|\mathbf{E}^\mathrm{H}(\tilde{f}_d)\mathbf{b}e^{-j\pi\tilde{\varepsilon} N_c{T_s}}\| = \|\mathbf{b}\|$ is a constant,~\eqref{eq:opt_3} can be further expressed as
\begin{equation}\label{eq:opt_4}
  \max_{\tilde{f}_d,\tilde{\varepsilon}} \left\|\Re[\mathbf{E}^\mathrm{H}(\tilde{f}_d)\mathbf{b}e^{-j\pi\tilde{\varepsilon} N_c{T_s}}]\right\|^2.
\end{equation}

Denote $\mathbf{u}=\mathbf{E}^\mathrm{H}(\tilde{f}_d)\mathbf{b}e^{-j\pi\tilde{\varepsilon} N_c{T_s}}$ as a complex vector. Since $\left\|\Re(\mathbf{u})\right\|^2 = \left\|\frac{\mathbf{u}+\mathbf{u}^*}{2}\right\|^2 = \frac{1}{4}(\mathbf{u}+\mathbf{u}^*)^\mathrm{H}(\mathbf{u}+\mathbf{u}^*) = \frac{1}{2}\mathbf{u}^\mathrm{H}\mathbf{u}+\frac{1}{2}\Re(\mathbf{u}^\mathrm{T}\mathbf{u}) = \frac{1}{2}\|\mathbf{b}\|^2+\frac{1}{2}\Re(\mathbf{u}^\mathrm{T}\mathbf{u})$, we can rewrite~\eqref{eq:opt_4} as
\begin{equation}\label{eq:opt_5}
  \max_{\tilde{f}_d,\tilde{\varepsilon}} \Re(\mathbf{u}^\mathrm{T}\mathbf{u}),
\end{equation}
which can be further written as
\begin{equation}\label{eq:opt_6}
  \max_{\tilde{f}_d,\tilde{\varepsilon}} \Re\{[\mathbf{b}^\mathrm{T}\mathbf{E}^\mathrm{*}(\tilde{f}_d)\mathbf{E}^\mathrm{H}(\tilde{f}_d)\mathbf{b}]e^{-j2\pi\tilde{\varepsilon} N_c{T_s}}\}.
\end{equation}

The optimization problem in~\eqref{eq:opt_6} can be decomposed into two parts which are associated with $\tilde{f}_d$ and $\tilde{\varepsilon}$, respectively. Finally, the optimal maximum Doppler shift $\bar{f}_d$ and the optimal OFO $\bar{\varepsilon}$ are estimated by
\begin{equation}\label{eq:fd}
  \bar{f}_d = \arg\max_{\tilde{f}_d} \|\mathbf{b}^\mathrm{T}\mathbf{E}^\mathrm{*}(\tilde{f}_d)\mathbf{E}^\mathrm{H}(\tilde{f}_d)\mathbf{b}\|,
\end{equation}
and
\begin{equation}
  e^{j2\pi\bar{\varepsilon}{N_c}{T_s}} = \frac{\mathbf{b}^\mathrm{T}\mathbf{E}^\mathrm{*}(\bar{f}_d)\mathbf{E}^\mathrm{H}(\bar{f}_d)\mathbf{b}} {\|\mathbf{b}^\mathrm{T}\mathbf{E}^\mathrm{*}(\bar{f}_d)\mathbf{E}^\mathrm{H}(\bar{f}_d)\mathbf{b}\|},
\end{equation}
respectively. Since only one variable is in the object function of~\eqref{eq:fd}, one dimensional searching is sufficient to obtain $\bar{f}_d$.

\vspace*{-0.0in}
\subsection{CFO Compensation and Channel Estimation}\label{ssec:chn_est}
Since each beamforming branch is affected by only one dominant CFO, we can perform CFO compensation directly. For the $i$th branch, the CFO can be estimated by $\bar{f}_d\cos{\theta_i}+\bar{\varepsilon}$. Then the $n$th time-domain sample in the $m$th OFDM block of the $i$th branch after CFO compensation can be easily given by
\begin{equation}
  \bar{r}_{m,n}(\theta_i) = e^{-j2\pi(\bar{f}_d\cos{\theta_i}+\bar{\varepsilon})(m{N_s}+n){T_s}} {r}_{m,n}(\theta_i).
\end{equation}

Now the channel for each branch after pre-processing can be considered as time-invariant. The conventional channel estimation methods can be carried out to estimate the channel response for each beamforming branch~\cite{Zhang11Multiple,Yang13Beamforming}. Then, the maximum-ratio-combining (MRC) is utilized to detect the transmitted data through $Q$ beamforming branches.

\vspace*{-0.0in}
\section{Simulation Results}
In this section, we evaluate the performance of the proposed transmission scheme through numerical simulations. We consider an OFDM system with $N_b=5$ blocks in a frame where the first block is the training block with two identical halves in time domain. Each block has $N_c=256$ subcarriers and the length of CP is $N_{cp}=32$. %while the maximum delay of channel is $16$.
The wavelength of carrier wave is $\lambda=0.1\mathrm{m}$ and the duration of each block is $T_b=T_sN_c=0.1\mathrm{ms}$. The speed of the train is $v=360\mathrm{km/h}$, which means that the normalized maximum DFO is $f_d T_b=0.1$. The normalized OFO which is defined as $\varepsilon T_b$ is randomly generated between -0.4 and 0.4. The mean square error (MSE) for the maximum Doppler shift is defined as $MSE=\frac{1}{N}\sum\nolimits_{n=1}^{N}{({{\bar{f}_d^n T_b}-f_dT_b})^2}$, where $N$ is the number of Monte Carlo trials and $\bar{f}_d^n$ is the $n$th estimation value of $f_d$. Similarly, we can define the MSE for the OFO. We assume the Jakes' channel model from the BS to the moving terminal as in~\cite{Zheng03Simulation}.
The antenna element spacing for the ULA is $d=0.45\lambda$ and the beamforming network is designed with the interval of $1^\circ$, that is $Q=181$. % for angles between $0^\circ$ and $180^\circ$ with the interval of $1^\circ$.
Some typical techniques in the literature are selected as benchmarks, which are
\begin{itemize}
\item The ML based OFO estimator proposed in \cite{Morelli00Carrier} when there is no Doppler spread, called MLE.
\item The joint maximum Doppler spread and OFO estimator proposed in \cite{Souden09Robust}. This method is based on covariance-matching and is denoted as COMAT.
\item The joint OFO and doubly-selective channel estimator proposed in \cite{Nguyen10Pilot}. The generalized complex-exponential (GCE) BEM is employed to characterize the time-varying channel and the ML technique is used in this scheme, thus we call it GCE-BEM-ML.
\end{itemize}

For fairness comparison, these schemes have been extended to the multi-antenna case at the receiver. In MLE, the OFO is jointly estimated among all the receive antennas. In COMAT, the received training blocks on all the antennas are exploited to estimate the covariance matrix more accurately. In GCE-BEM-ML, the OFO and BEM coefficients are jointly estimated with received signals on all antennas at first, then the transmitted signal is detected via MRC on all antennas. We consider two frame structures in GCE-BEM-ML: the first structure dedicates first block as the training block while the second one considers both the first and last blocks as training blocks. They are called GCE-BEM-ML-I and GCE-BEM-ML-II, respectively.
Moreover, the maximum DFO and the OFO in these works are selected the same as that in our scheme.

\begin{figure}[t]
\centering
\includegraphics[scale=0.45]{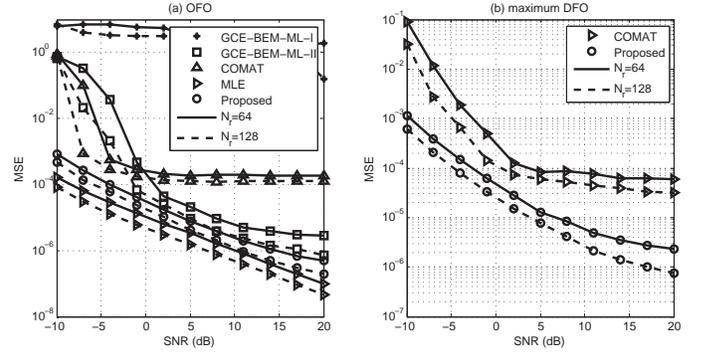}
\vspace*{-0.2in}
\caption{MSE performance of the CFOs estimation: (a) OFO; (b) maximum DFO.}
\label{fig:MSE_CFO}
\vspace*{-0.0in}
\end{figure}

We first evaluate the performance of the CFOs estimation method proposed in Section~\ref{ssec:fo_est}. The MSE performance of the OFO and the maximum DFO is plotted versus SNR under different numbers of antennas, $N_r$, in Figure~\ref{fig:MSE_CFO}(a) and \ref{fig:MSE_CFO}(b), respectively. %The solid and dashed curves correspond to $N_r=64$ and $N_r=128$ antennas, respectively.
We see that the proposed method performs better than COMAT, GCE-BEM-ML-I and GCE-BEM-ML-II. With the increase of $N_r$, the MSE reduces significantly since more refined beams are achieved. As introduced in \cite{Morelli00Carrier}, MLE achieves the Cramer-Rao bound (CRB) for OFO estimation when there is no Doppler spread, thus the performance of MLE can be considered as the upper bound for OFO estimation.

\begin{figure}[t]
\centering
\includegraphics[scale=0.47]{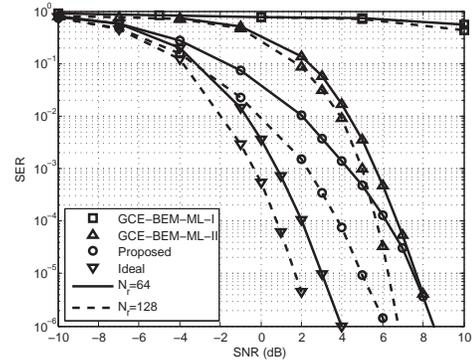}
\vspace*{-0.0in}
\caption{SER performance of the proposed scheme.}
\label{fig:SER_SNR}
\vspace*{-0.0in}
\end{figure}

In Figure~\ref{fig:SER_SNR}, we further present the symbol error rate (SER) performance of the proposed scheme under different values of $N_r$. For comparison, GCE-BEM-ML and the ideal transmission scheme are also included.
In the ideal transmission scheme, we assume none CFOs at the receiver and the symbols on each subcarrier are recovered via MRC on all antennas.
From this figure, GCE-BEM-ML-I has poor SER performance when SNR changes from $-10\mathrm{dB}$ to $10\mathrm{dB}$, while GCE-BEM-ML-II benefits from the multi-antenna diversity gain with increased training blocks. The proposed scheme outperforms GCE-BEM-ML-I and GCE-BEM-ML-II dramatically. Compared with GCE-BEM-ML-II,
i) The proposed scheme has better SER performance while reducing the resource consumption of training blocks used in channel estimation, e.g. one training block in the proposed scheme and two training blocks in GCE-BEM-ML-II; ii) The proposed scheme simplifies the receiver design since the channel for each beamforming branch after CFO compensation can be considered as time-invariant.

The SER performance of the ideal transmission scheme also demonstrates the effectiveness of the proposed scheme. Moreover, the SER can be further reduced if more antennas are configured at the receiver. Therefore, the simulation results indicate that the proposed scheme can significantly overcome the SER degradation caused by the time-varying multipath channel with multiple CFOs.

\vspace*{-0.0in}
\section{Conclusion}
In this correspondence, the high-mobility OFDM transmission with multiple DFOs and single OFO has been considered. With the large-scale antenna array, the time-varying multipath channel can be divided into multiple channels with single CFO in spatial domain. Then the conventional CFO compensation and channel estimation methods can be directly used to recover the transmitted data. Simulation results show that the proposed scheme is effective under high-mobility scenarios.

\bibliographystyle{IEEEtran}
\bibliography{References}

% that's all folks
\end{document}